# Single-Defect Spectroscopy via Random Telegraph Noise in Graphene-Contacted ReS$_2$-hBN Heterostructures

Shubhrasish Mukherjee[1,#]*, Gaurab Samanta[1,$], Shubhadip Moulick[1,$$], Ruta Kulkarni[2], Kenji Watanabe[3], Takashi Taniguchi[4], Arumugum Thamizhavel[2], and Atindra Nath Pal[1]*

[1]*S. N. Bose National Center for Basic Science, Sector III, Block JD, Salt Lake, Kolkata – 700106*

[2]*Department of Condensed Matter Physics and Material Science, Tata Institute of Fundamental Research, Mumbai 400 005, India*

[3]*Research Center for Electronic and Optical Materials, National Institute for Materials Science, 1-1 Namiki, Tsukuba 305-0044, Japan*

[4]*Research Center for Materials Nano architectonics, National Institute for Materials Science, 1-1 Namiki, Tsukuba 305-0044, Japan*

Email: shubhraphysicsbu@gmail.com , atin@bose.res.in

[#] Current Affiliation: Institute for Functional Intelligent Materials, National University of Singapore, 117544, Singapore.

[$] Current Affiliation: Institut de Physique et Chimie des Matériaux de Strasbourg (IPCMS), CNRS, Université de Strasbourg, 67034 Strasbourg, France.

[$$] Current Affiliation: Applied physics department, Aalto University, Finland.

**Abstract**

Defect spectroscopy in two-dimensional (2D) field-effect transistors (FETs) requires device architectures that suppress contact and disorder artifacts while preserving intrinsic carrier dynamics. Here, we demonstrate ReS$_2$-hBN FETs with few-layer graphene (FLG) van der Waals contacts that form nearly barrier-free interfaces, enabling intrinsic transport in ReS$_2$, an anisotropic, low-symmetry TMDC rarely exhibiting disorder-free behavior. The clean ReS$_2$-FLG platform allows direct observation of random telegraph noise (RTN) even in micron-scale channels, manifested as discrete two-level current fluctuations between 90-150 K arising from stochastic trapping at localized hBN defect sites. With increasing temperature, the RTN evolves into a 1/f spectrum as multiple traps activate. Statistical analysis of RTN amplitudes and capture-emission kinetics identifies substitutional carbon-related centers in hBN as dominant defects. These findings establish a generalizable approach for probing dielectric-origin defect dynamics in intrinsically conducting, low-symmetry 2D semiconductors.

## Introduction

Understanding charge trapping and defect dynamics is essential for advancing two-dimensional (2D) field-effect transistors (FETs) toward reliable nanoelectronics, optoelectronics and quantum technologies[1–4]. In atomically thin semiconductors, even a single defect can modulate local electrostatics, alter carrier mobility, and induce stochastic conductance fluctuations that obscure intrinsic transport properties[5]. Reliable defect spectroscopy therefore demands device architectures that minimize extrinsic artifacts while preserving the intrinsic carrier response of the 2D channel[1,6,7]. A major limitation is the contact interface: conventional metal–semiconductor contacts can introduce Schottky barriers, Fermi-level pinning, and excess low-frequency noise that mask subtle defect-driven conductance fluctuations[8,9]. In contrast, semi-metallic van der Waals contact engineering offers a versatile approach to lower injection barriers and suppress parasitic noise, thereby providing a cleaner transport baseline for defect spectroscopy without invoking interfacial chemistries that risk perturbing the 2D channel[10–12].

Random telegraph noise (RTN), discrete two-level current fluctuations from stochastic carrier capture and emission, serves as a direct fingerprint of individual defects[5,13]. It represents both a challenge, by limiting reliability in nanoscale electronic devices and degrading coherence in semiconductor-based quantum architectures[7], and an opportunity, as a natural source of true randomness for hardware random-number generation[14]. Within 2D semiconductors, RTN has been established primarily in $MoS_2$ devices, including those with conventional metal contacts, and also in vdW-assembled heterostructures[15–18]. Crystalline hexagonal boron nitride (hBN) as dielectric substrate can suppress environmental disorders yet introduces its own intrinsic defect landscape that governs trapping and low-frequency noise[19,20]. This underscores the need for contact-clean platforms that can resolve dielectric-origin defects in operating devices.

Here, we realize $ReS_2$-hBN FETs with graphene-based van der Waals contacts as a sensitive platform for defect spectroscopy in a group-VII TMDC. Clear two-level RTN is observed between 90–150 K, arising from stochastic charge exchange between the $ReS_2$ channel and defect states in proximate to the hBN. Statistical analyses quantify defect densities, capture–emission kinetics, and energetic levels, consistent with substitutional carbon–related centers in hBN; at T > 150 K, the spectrum evolves to $1/f$ as multiple traps activate. By extending RTN studies beyond MoS$_2$ and directly resolving dielectric-origin defects in ReS$_2$, this work establishes vdW contact engineering as an effective route to interrogate, and ultimately engineer, microscopic defect landscapes in 2D heterostructures for reliable, CMOS-compatible nanoelectronics.

## Experimental results

2D van der Waals devices were fabricated using mechanically exfoliated few-layer ReS$_2$ as the channel, with few-layer graphene electrodes on an hBN substrate. Fabrication details are provided in the **Methods section**. A schematic and optical image of a representative device are shown in **Figure 1a**. Transport and noise measurements were performed on several devices with both metallic (Ti/Au and Cr/Au) contacts and graphene contacts (see our previous work[12]). In this paper, we concentrate on the results obtained for one of the graphene contacted devices (RG1). **Figure 1b** shows the Raman spectrum of the ReS$_2$ flake, displaying all fundamental modes (A$_{1g}$, E$_{1g}$, and E$_{2g}$)[21]. Room-temperature transport measurements (**Figure S1**) confirm linear $I_{ds} - V_{ds}$ characteristics at all gate biases, consistent with ohmic-like graphene–ReS$_2$ contacts reported earlier[12]. **Figure 1c** shows the semilogarithmic transfer characteristics ($I_{ds} - V_{bg}$) at $V_{ds} = 50\ mV$ for various temperatures (see **Figure S2a** for linear-scale plots). Irrespective of temperatures, the device offers n-type conduction with excellent ON/OFF ratio $> 10^4$. Furthermore, below $235\ K$, the device exhibits insulating behaviour at all $V_{bg}$, whereas a mild metal-like behaviour is found at $T > 235\ K$ for higher $V_{bg}$ ($> 35\ V$). Thermal activation of the charge carriers primarily regulates the electrical transport of the device in that temperature range ($T < 235\ K$), as seen by the increase in conductivity with increasing temperature[22]. The threshold voltage ($V_{th}$) was estimated by extrapolating the linear portion of the transfer characteristics, and it decreases markedly with increasing temperature, reaching $\sim 0\ V$ near room temperature (**Figure S2b**). **Figure S2c** presents the Arrhenius plot of conductivity ($\sigma$) at different gate voltages. The subthreshold swing (SS) was extracted from the transfer characteristics using the relation[23]:

$$SS = \left| \left( \frac{d\ (\log(I_{ds}))}{dV_{bg}} \right)^{-1} \right| \dots\dots\dots\dots\dots\dots\dots\dots\dots (1.1)$$

At room temperature, the device exhibits SS of $\sim 3.5\ V/dec$ (see **Supplementary Note 2** for details). This value is consistent with previous reports on ReS$_2$ FETs and reflects a combination of weak gate electrostatics in back-gated geometries and interfacial trap states[24]. **Figure 1d** depicts the temperature-dependent field effect mobility (μ) which is determined by using the realtion[25]

$$\mu = \frac{L}{WC_{total}} \frac{1}{V_{ds}} \frac{dI_{ds}}{dV_{bg}} \quad \ldots\ldots\ldots\ldots\ldots\ldots\ldots\ldots\ldots\ldots (1.2)$$

where, L and W are the channel length and width respectively, $C_{total}$ is the total capacitance which is basically a parallel combination of $285\ nm$ SiO$_2$ and $\sim 10\ nm$ of hBN. Here, it is observed that, μ initially increased from $\sim 21\ cm^2V^{-1}s^{-1}$ at $88\ K$ to $\sim 30\ cm^2V^{-1}s^{-1}$ at $130\ K$ and then decreases with increasing of temperature. Such an anomalous behaviour of μ can be explained by the different scattering mechanisms like coulomb scattering, acoustic and optical phonon scattering, interfacial phonon scattering etc[26,27]. At higher temperature ($T >$ 130 $K$), the electron-phonon scattering dominates the electron mobility and can be scaled as, $\mu \sim T^{-\nu}$, where $\nu$ is the exponent that depends on the dominant phonon scattering mechanism[28] and is determined to be 1.38. On the other hand, at low temperature, charge impurities located randomly at the sample, interfaces or substrate controls the electron transport and effectively increases $\mu$ with increasing temperature. Following the Matthiessen's rule, the temperature dependent $\mu$ can be fitted by[29]

$$\frac{1}{\mu} = \frac{1}{M_p T^{-\frac{3}{2}}} + \frac{1}{M_i T^{\frac{3}{2}}} \quad \ldots\ldots\ldots\ldots\ldots\ldots\ldots\ldots\ldots\ldots\ldots\ldots (1.3)$$

where, $M_p$ and $M_i$ are the relative contributions of phonon mediated and impurity scattering respectively.

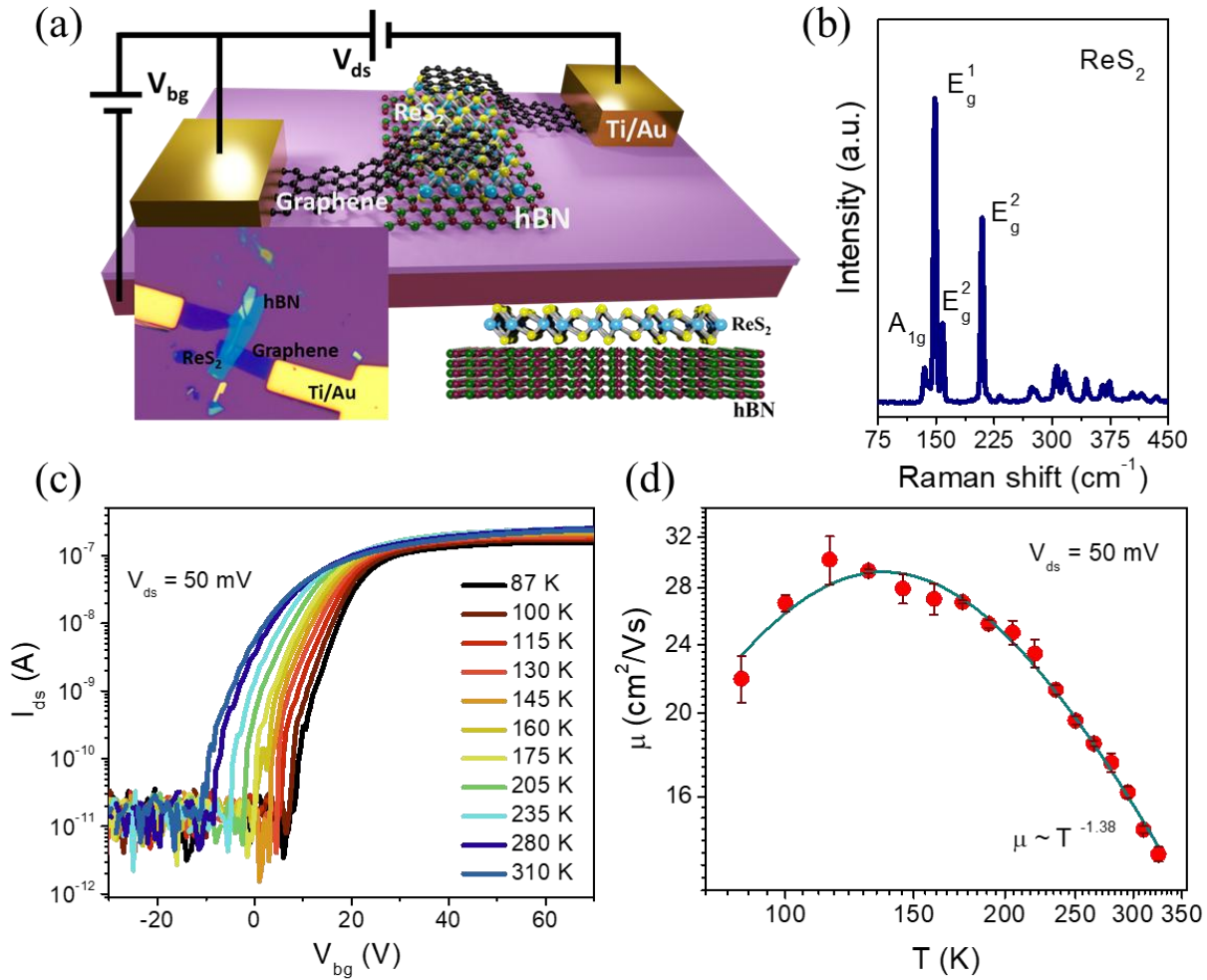

*Figure 1: Electrical transport of ReS$_2$-hBN hybrid device: (a) Schematic of device. Insets show the optical image and the vertical cryosection of the hybrid interface. (b) Raman spectra of the exfoliated ReS$_2$ layer. (c) Temperature dependent $I_{ds} - V_{bg}$ characteristics of the device. (d) Temperature dependent field effect mobility (μ) of the device.*

To gain deeper insight about the charge-carrier dynamics at the surface and interfaces of this hybrid ReS$_2$-hBN system, we employed a two-probe *ac* digital-signal-processing approach to study low-frequency conductance fluctuations (current noise spectroscopy) over the temperature range of $87 - 310\ K$[30,31]. The schematic of the noise measurement setup and a representative $1/f$ noise spectrum with a flat thermal noise background are shown in **Figure S3.** Furthermore, the noise power spectral density (PSD) at a fixed $V_{bg}$ scales as $S_I \propto I_{ds}^2$, indicating that the $1/f$ noise originates from intrinsic conductance fluctuations within the channel rather than from contact or injection-related processes (**Figure S4**). **Figure 2a** presents the drain-current ($I_{ds}$) time series at different temperatures under constant bias conditions ($V_{ds} = 50\ mV\ and\ V_{bg} = 65\ V$), where distinct step-like features are observed below $150\ K$. The temporal current fluctuations in $I_{ds}$ arose from the random capture and emission of charge carriers. These fluctuations between two well-defined current levels are attributed to a single type of defect state and are commonly referred to as random telegraph signals (RTS) or two-level fluctuations (TLF)[7,32] (**Figure S5**). As expected, the RTS amplitude is strongest at low temperatures, gradually weakens with increasing temperature, and eventually vanishes near room temperature. The corresponding power spectral density (PSD) in **Figure 2b** reveals random telegraph noise (RTN), which deviates markedly from the $1/f$ dependence (dotted line) over the temperature range where RTS is observed. This deviation is further highlighted in **Figure 2c** by the frequency-normalized PSD ($fS_I$), expected to be constant for pure $1/f$ noise. Meanwhile, the PSD can be well fitted by a Lorentzian function[33] i.e.,

$$\frac{S_I(f)}{I_{ds}^2} = \frac{A}{f} + \frac{Bf_c}{f^2 + f_c^2} \quad \ldots\ldots\ldots\ldots\ldots\ldots\ldots\ldots\ldots\ldots (1.4)$$

where $f_c = \frac{1}{\tau}$ is the characteristic transition frequency of RTS, $\tau$ the characteristic time, and $A$ and $B$ the relative contributions of $1/f$ noise and RTN. As shown in **Figure 2b**, $f_c$ shifts with temperature from a few mHz to a few Hz. Since carrier capture cross-section is thermally activated,

$$f_c = f_0 e^{-\frac{E_a}{K_B T}} \quad \dots\dots\dots\dots\dots\dots\dots\dots\dots\dots\dots\dots (1.5)$$

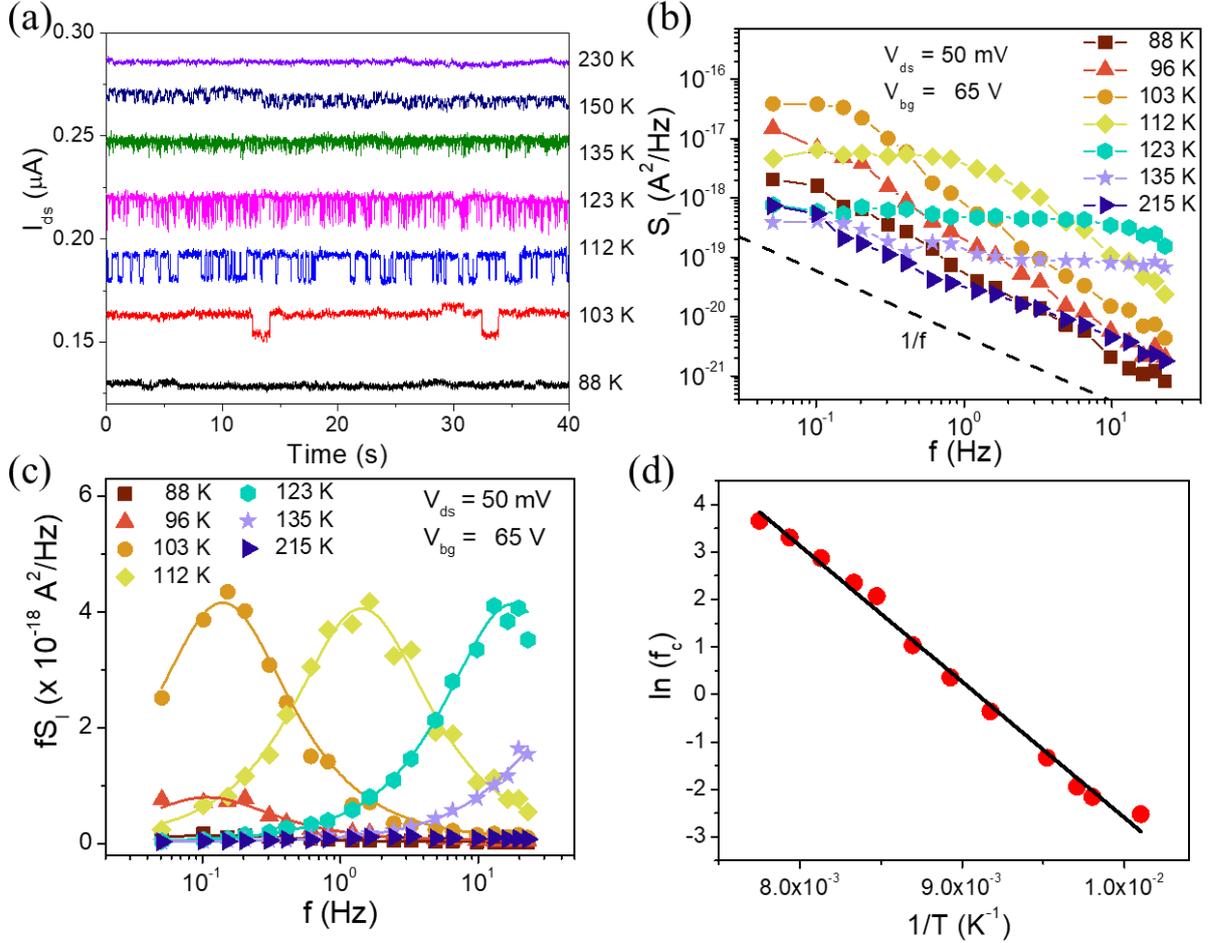

*Figure 2: Observation of random telegraph noise in ReS$_2$-hBN hybrid device: (a) Temperature-dependent (a) time series of source-drain current ($I_{ds}$), and (b) corresponding power spectral densities ($S_I$) of the device. The black dashed lines represent the 1/f reference. (c) Plot of $fS_I$ at different temperatures; the solid lines represent Lorentzian fits to the characteristic peaks. (d) Arrhenius plot of ln($f_c$) as a function of 1/T; the black line represents a linear fit. All measurements were performed under the same experimental conditions of $V_{ds}$ = 50 mV, $V_{bg}$ = 65 V.*

where $k_B$ is the Boltzmann constant and $E_a$ is the trap activation energy. An Arrhenius plot of $\ln(f_c)$ versus $\frac{1}{T}$ (**Figure 2d**) yields $E_a \sim 208\ meV$, consistent with values reported for other TMDC FETs[22,34]. Notably, $f_c$ is nearly independent of $V_{ds}$, and the bias-dependent $fS_I$ and $f_c$

(**Figure S6**) confirms that trap kinetics are primarily governed by vertical electrostatics rather than lateral bias fields[22].

To probe the defect energetics, we examined the gate-voltage dependence of the two-level fluctuations, since $V_{bg}$ shifts the channel Fermi level relative to trap states. **Figure 3a** shows the $I_{ds}$ time series at different $V_{bg}$ ($V_{ds} = 50\ mV, T = 110\ K$), where the dwell times in the low- and high-current states define the as emission ($\tau_e$) and capture ($\tau_c$) times (**Figure S5c** and **Figure S7**). Current histograms (**Figure 3b**) exhibit two Gaussian peaks that merge near $150\ K$ as additional traps activate (**Figure S8**). Statistical analysis[16] yields average $\tau_e$ and $\tau_c$, whose PDFs are shown in **Figures 3c,d** and gate dependence in **Figure 3e** (see **Figure S10** for other T). While $\tau_c$ decreases with $V_{bg}$, $\tau_e$ remains nearly constant, indicating that the traps are located in hBN rather than in ReS$_2$, since capture is modulated by electrostatics whereas emission is bias-insensitive[15,35]. The temperature-dependent lifetimes (**Figure S11**) exhibit Arrhenius behaviour, indicating thermally activated kinetics and enabling extraction of the defect energy level ($E_T$) relative to the channel Fermi level ($E_F$) from $\tau_c/\tau_e \propto \exp((E_T - E_F)/k_B T)$. **Figure 3f** represents ($E_T - E_F$) as a function of $V_{bg}$. The uncertainty in determining $E_F$ makes it impossible to pinpoint the precise $E_T$ level, but this method can provide an indication of the relative position of $E_T$ from $E_F$. Also, the physical location ($d$) of the defect sites can be determined by the following equation[15]

$$d \sim - t_{sub} \frac{K_B T}{q} \frac{\delta \ln\left(\frac{\tau_c}{\tau_e}\right)}{\delta V_{bg}} \quad \ldots\ldots\ldots\ldots\ldots\ldots\ldots\ldots\ldots\ldots (1.6)$$

where, $t_{sub} = t_{hBN} + t_{SiO_2}$ is the total thickness of the substrate. Here, we found $d \sim 3\ nm$ from the ReS$_2$-hBN interface. Using this equation, the defect level is found to be inside hBN and $\sim 3\ nm$ below the interface. **Figure 3g** represents the corresponding current noise ($S_I$) spectra, where clear Lorentzian profile $\left(\frac{1}{f^2}\right)$ is observed irrespective of the gate voltages (**Figure S9**). The carrier capture and emission by a defect state are considered as an elastic tunnelling event manifesting as RTS in the time series and a Lorentzian spectrum in the frequency domain.

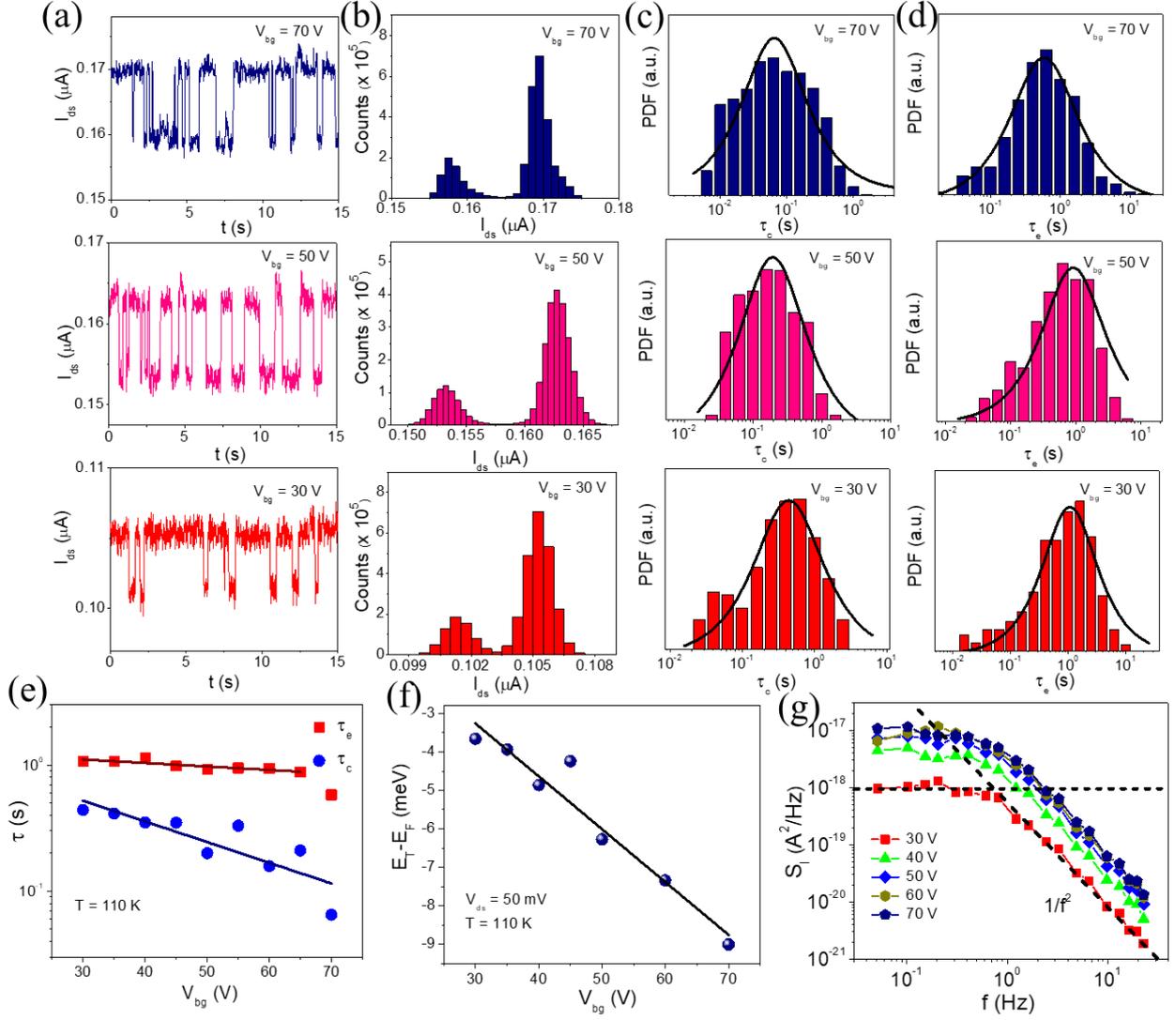

*Figure 3: Gate voltage dependent RTS in ReS$_2$-hBN hybrid device*: (a) Time series of source-drain current ($I_{ds}$), histograms of (b) $I_{ds}$, (c) capture ($\tau_c$) and (d) emission ($\tau_e$) times, respectively; for three different gate voltages: 70 V, 50 V and 30 V (top to bottom panels). The black lines represent Gaussian fits used to extract the average characteristic times. (e) Average characteristics times ($\tau_c$ and $\tau_e$) as a function of back gate voltage. (f) Relative energetic position of the defect with respect to the Fermi level ($E_T$-$E_F$) of ReS$_2$ in the hybrid device. (g) Corresponding low frequency noise spectra of the device under identical experimental conditions ($V_{ds}$ = 50 mV, T = 110 K).

To understand the relative variance of the noise, the PSD ($S_I$) is integrated over the measured frequency bandwidth and can be represented as

$$\frac{<\delta g^2>}{<g>^2} = \frac{<\delta I_{ds}^2>}{<I_{ds}>^2} = \frac{1}{I_{ds}^2} \int_{f_1}^{f_2} S_I(f)df \quad \dots\dots\dots\dots\dots\dots (1.7)$$

The temperature dependent conductance fluctuation $\left(\frac{<\delta g^2>}{<g>^2}\right)$ is represented in **Figure 4a**. The relative variance increases by more than an order of magnitude in the temperature range of $88 - 150\ K$, coinciding with the observed RTS. This clearly indicates that the enhanced noise within this temperature window (~ 88–150 K) originates entirely from the system's two-level conductance fluctuations (RTN), as confirmed when the total integrated noise is decomposed into two components: $1/f$ and RTN $\left(\frac{1}{f^2}\right)$. The gate-dependent noise data provide direct insight into the underlying noise mechanism. With increasing $V_{bg}$ (or carrier density), the normalized noise magnitude follows $S_I/I_{ds}^2 \propto 1/\Delta V_{bg}^2$ (**Figure S12**), indicating that the dominant contribution arises from carrier number fluctuations. This behaviour is consistent with charge exchange between the ReS$_2$ channel and interfacial trap states at the ReS$_2$-hBN interface, as described by the McWhorter model[36,37]. The corresponding interfacial defect density can be estimated using the $1/f$ noise as[36,38]

$$D_{it} = f S_I(f) <R^2> \frac{WLC^2}{e^2 K_B T} \quad \dots\dots\dots\dots\dots\dots\dots\dots\dots\dots\ (1.8)$$

Where, $D_{it}$ is the defect density, $R$ is the channel resistance, $W$ is the width, $L$ is the length of the device etc. **Figure 4b** shows the variation of $fS_I$ as a function of temperature, exhibiting a linear dependence consistent with Equation (1.8). The extracted defect density is estimated to be on the order of $\sim 4 \times 10^{11}\ cm^{-2}eV^{-1}$, which is comparable to the previously reported values for TMDC devices[17,39] (see **Supplementary note 9** for details). To quantify the noise magnitude, we have calculated phenomenological Hooge parameter ($\gamma_H$) from the empirical relation[40]

$$\frac{S_I}{I_{ds}^2} = \frac{\gamma_H}{Nf^\alpha} \quad \dots\dots\dots\dots\dots\dots\dots\dots\dots\dots\dots\dots\ (1.9)$$

where, $N$ is the total number of carriers in the channel. The calculated $\gamma_H$ becomes as low as $\sim 1.6 \times 10^{-2}$ which is lower or comparable to the reported values[41–43] (see **Supplementary Note 10** and **Figure S13** for details). To gain a deeper physical understanding, **Figure 4c** presents a schematic of the ReS$_2$-hBN heterostructure illustrating the possible carbon-related defects in the underlying hBN layers. As reported in previous theoretical and experimental studies, various types of intrinsic and extrinsic defects can form within the hBN lattice, primarily determined by growth conditions or post-growth processing[19,35,44]. Common defects in high-quality single-crystal hBN (where no intentional doping is introduced) include native point defects such as nitrogen vacancies ($V_N$) and boron vacancies ($V_B$), as well as carbon-related impurity.[35,45] These impurities act as a defect potential which relaxes and deforms locally as a result of the charge transfer between the channel and the defect location. Consequently, the observed TLF process can be modelled by an asymmetric two-state system separated by an energy barrier $E_a$ (**Figure 4d**)[Ref. https://journals.aps.org/prb/pdf/10.1103/PhysRevB.15.989, https://journals.aps.org/prl/pdf/10.1103/PhysRevLett.55.859 ]. Thermal activation enables random switching between these states, giving rise to the characteristic RTN. Depending on the trap location and gate bias, charge carriers may be captured from or emitted into the channel. The gate-voltage-dependent RTS switching behaviour can be interpreted using the band diagram shown in **Figure 4e**, where the effect of a small change in gate voltage ($\delta V_{BG}$) is indicated by the dotted line. In addition to the typical two-level RTS associated with a single defect, the device occasionally exhibits more complex behaviour with three discrete current levels (**Figure S14**). This so-called anomalous RTS has been previously attributed[46] to the involvement of multiple metastable defect configurations in addition to a single active trap, warranting further investigation to fully elucidate their physical origin.

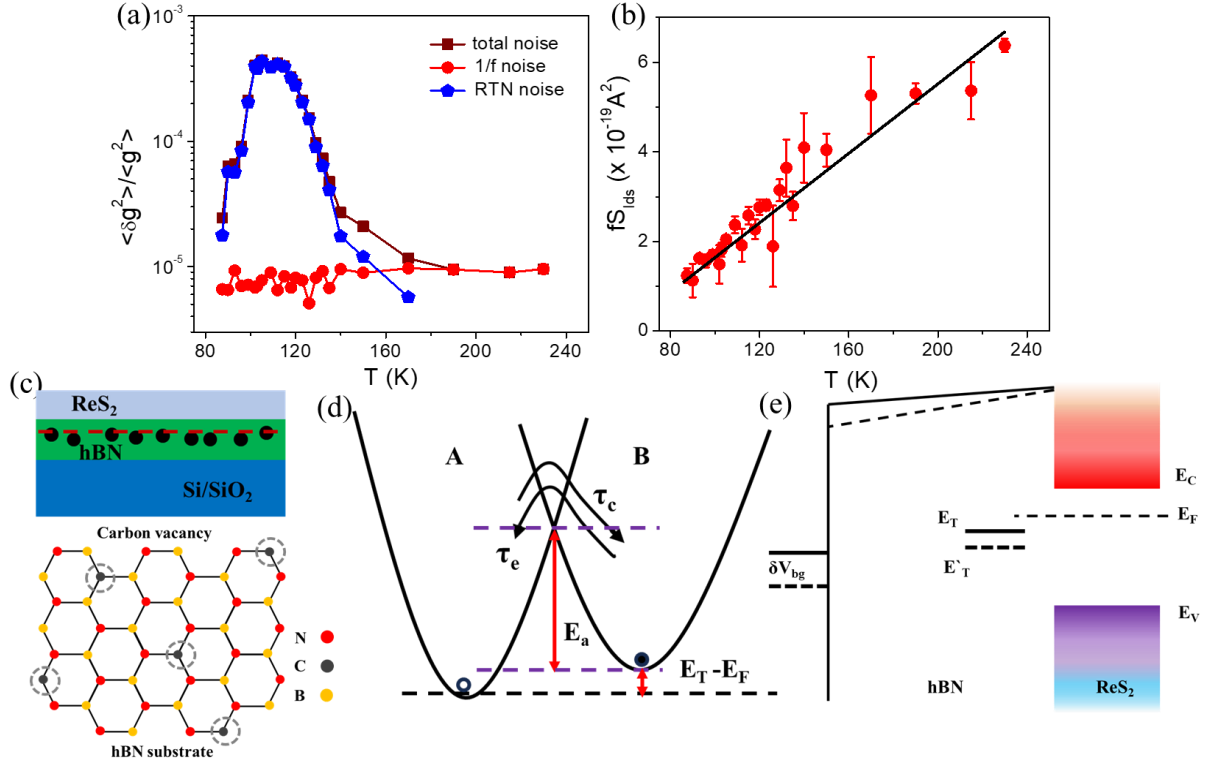

*Figure 4: Temperature dependence of low frequency noise:* *(a) Variation of normalized total conductance fluctuations as a function of temperature. The 1/f noise and RTN components are separated and shown individually. (b) Plot of $fS_I$ as a function of temperature. The solid black line represents the linear fit to the data. (c) Illustration of defects in the bottom hBN layer, with a close-up showing carbon defects substituting nitrogen atoms in hBN. (d) Schematic representation of an asymmetric double-well model describing the fluctuation process. (e) Energy band diagram depicting the energetic alignment between the trap level ($E_T$) and the Fermi level ($E_F$) at the ReS$_2$-hBN interface.*

## Conclusions

In summary, we have explored single-defect dynamics and low-frequency noise in large-area (~10 μm$^2$) ReS$_2$-hBN heterostructures with graphene van der Waals contacts, a configuration that minimizes contact-induced artifacts and enables intrinsic defect spectroscopy. Temperature- and gate-dependent transport and noise analyses reveal discrete two-level fluctuations arising from stochastic charge capture and emission at hBN defect sites. Statistical evaluation of capture-emission kinetics yields activation energies and energy levels consistent with substitutional carbon-related center located a few nanometres below the ReS$_2$-hBN interface. The power spectral density exhibits a nonmonotonic temperature dependence and a transition from $1/f$ to Lorentzian behaviour, signifying a crossover from distributed to discrete defect kinetics. The extracted interfacial defect density (~ $4 \times 10^{11}$ cm$^{-2}$ eV$^{-1}$) and activation energy (~208 meV) are consistent with intrinsic hBN defect states reported previously, corroborating their dominant role in charge trapping. Collectively, these results provide direct microscopic insight into the dielectric-origin defect landscape of hBN and its coupling to 2D semiconductors, advancing the understanding of noise mechanisms in van der Waals nanoelectronics. Beyond elucidating fundamental defect–transport interplay, this work establishes graphene-contacted ReS$_2$-hBN devices as a robust platform for probing single-defect physics and guiding defect-engineered growth strategies for atomically thin electronic, optoelectronic, and quantum devices.

## Methods

**Synthesis of ReS$_2$ single crystal:** A two-step process has been adopted for the growth of ReS$_2$ Single crystals. High purity Re and sulfur powders were thoroughly mixed in an agate mortar and cold pressed into a pellet. This pellet was then loaded into a thick-walled quartz ampoule and sealed in a vacuum of ~ 10$^{-6}$ Torr. The quartz ampoule was then heated in a box type furnace by slowly ramping the furnace to 1150°C at the rate of 15°C/hr. and held at this temperature for about 72 hrs. before cooling it to room temperature. This pre-reacted material was then subsequently used for chemical vapor transport. About 1 gm of the pre-reacted $ReS_2$ powder along with $I_2$ (150 mg) was taken in a 25 cm long quartz tube which was sealed in a vacuum of about $10^{-6}$ Torr. The quartz tube was placed in a two-zone furnace. The hot zone was maintained at 1000°C and the cold zone was maintained at 960°C and this gradient was maintained for about 8 days to enable the vapor transport and recrystallization of the material. Several nucleation happened and tiny flakes of $ReS_2$ single crystals crystallized at the cold end of the crucible.

**Device fabrication:** Graphene (supplied by SPI Supplies), layered hBN (supplied from NIMS, Japan), and few layered $ReS_2$ flakes, are mechanically exfoliated from their bulk crystals by using scotch tape on $Si/SiO_2$ substrate. Then, the different hybrid stacks (graphene-$ReS_2$-hBN) are fabricated by standard pick up and attach based dry transfer method. The few-layered hBN is a wide band gap semiconductor, which acts as the bottom insulating surface with fewer defects than $SiO_2$ and also provides an atomically flat platform for $ReS_2$ channels. The few layered graphene is put on top of $ReS_2$ which acts as the contact electrodes. The electrical contacts are fabricated by laser writer lithography (LW405D-MICROTECH) technique followed by metallization with e-beam evaporation of Ti-Au ($5 - 50\ nm$) with a well-controlled deposition rate (> 0.4 Å/s) at $10^{-6}$ mbar. The channel length of all the devices are $\sim 7\ \mu m$. All the devices are annealed at 300°C in the presence of Ar-$H_2$ gas for 3 hrs in order to get better electrical contacts.

**Materials and devices characterizations:** All exfoliated 2D materials were characterized by Raman spectroscopy (LabRAM HR Evolution, HORIBA, France) using a 532 nm excitation laser. Electrical transport and low-frequency noise measurements were carried out under high vacuum (>$10^{-5}$ mbar) in a custom-built cryogenic probe insert. Temperature control and measurement were performed using a Lake Shore temperature controller (Model 336). A Zurich Instruments MFLI lock-in amplifier, a Keithley 2450 source meter, and a Femto DLPCA-200 current-to-voltage preamplifier were employed for all experiments in an AC two-probe configuration with a carrier frequency of 83.67 Hz.

# Acknowledgements

This research has made use of the Thematic Unit of Excellence on Nanodevice Technology (grant no. SR/NM/NS-09/2011) and the Technical Research Centre (TRC) Instrument facilities of S.N. Bose National Centre for Basic Sciences, established under the TRC project of Department of Science and Technology (DST), Govt. of India. A.N.P. acknowledges DST Nano Mission: DST/NM/TUE/QM-10/2019. K.W. and T.T. acknowledge support from the JSPS KAKENHI (Grant Numbers 21H05233 and 23H02052) and World Premier International Research Center Initiative (WPI), MEXT, Japan.

## Author contributions

S.M. and A.N.P. conceived the project. S.M. designed the experiments. S.M. and G.S. fabricated the devices and performed the detailed transport measurements. S.Mo. helped to set up the low temperature measurements. S.M. performed the complete data analysis. R.K. and A.T. have grown and characterized the $ReS_2$ single crystal. K.W. and T.T. have grown and supplied the hBN single crystal. S.M. wrote the original manuscript. A.N.P. supervised the project, validated the analysis, and reviewed and edited the manuscript. All authors have read and approved the final version of the manuscript.

**Supporting Information**

# Single-Defect Spectroscopy via Random Telegraph Noise in Graphene-Contacted ReS₂-hBN Heterostructures

Shubhrasish Mukherjee[1,#]*, Gaurab Samanta[1,$], Shubhadip Moulick[1,$$], Ruta Kulkarni[2], Kenji Watanabe[3], Takashi Taniguchi[4], Arumugum Thamizhavel[2], and Atindra Nath Pal[1]*

[1]S. N. Bose National Center for Basic Science, Sector III, Block JD, Salt Lake, Kolkata – 700106

[2]Department of Condensed Matter Physics and Material Science, Tata Institute of Fundamental Research, Mumbai 400 005, India

[3]Research Center for Electronic and Optical Materials, National Institute for Materials Science, 1-1 Namiki, Tsukuba 305-0044, Japan

[4]Research Center for Materials Nano architectonics, National Institute for Materials Science, 1-1 Namiki, Tsukuba 305-0044, Japan

Email: shubhraphysicsbu@gmail.com , atin@bose.res.in

[#] Current Affiliation: Institute for Functional Intelligent Materials, National University of Singapore, 117544, Singapore.

[$] Current Affiliation: Institut de Physique et Chimie des Matériaux de Strasbourg (IPCMS), CNRS, Université de Strasbourg, 67034 Strasbourg, France.

[$$] Current Affiliation: Applied physics department, Aalto University, Finland.

**Supplementary Note 1:** Room temperature transport characteristics of the hybrid device

**Supplementary Note 2:** Low temperature transport characteristics and calculation of interfacial defect density from subthreshold swing ($SS$)

**Supplementary Note 3:** Low frequency noise measurement technique

**Supplementary Note 4:** Drain-source bias ($V_{ds}$) dependent noise in room temperature:

**Supplementary Note 5:** Observation of $1/f$ and $1/f^2$ noise spectra

**Supplementary Note 6:** Drain-source bias ($V_{ds}$) dependent noise in low temperature

**Supplementary Note 7:** Evaluation of time series, histogram, and noise PSD of ($I_{ds}$) at different gate voltages and temperatures

**Supplementary Note 8:** Evaluation of characteristics time constants at different gate voltages and temperatures

**Supplementary Note 9:** Gate dependence and microscopic mechanism of low frequency noise

**Supplementary Note 10:** Low frequency $1/f$ noise in additional device and comparison of noise figure of merit

**Supplementary Note 11:** Observation of anamolous RTS

1. **Room temperature transport characteristics of the hybrid device:**

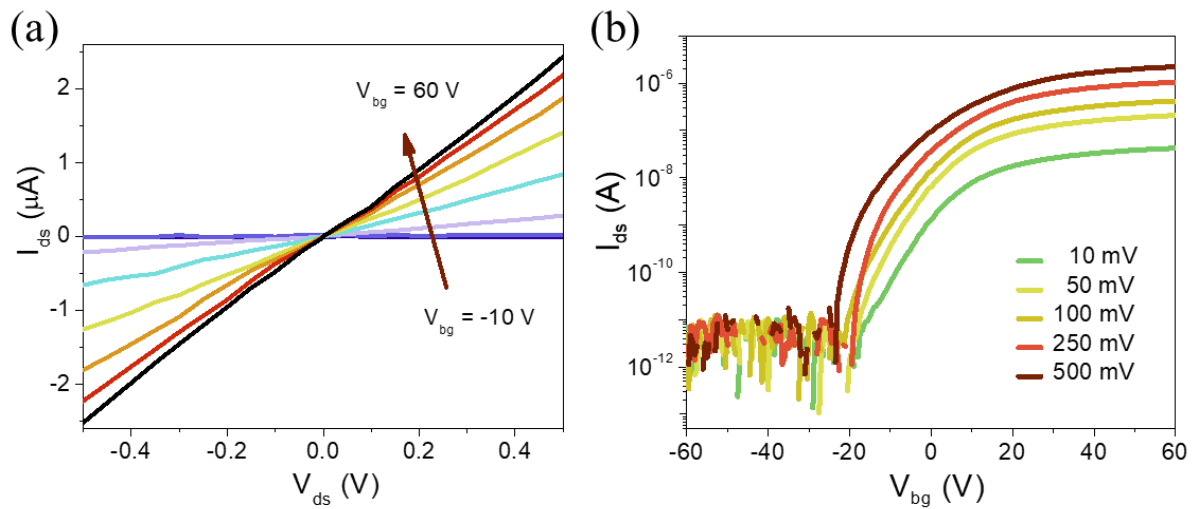

*Figure S1: Room temperature characteristics of the fabricated device: (a) $I_{ds}$-$V_{ds}$ characteristics at different gate voltages ($V_{bg}$). (b) $I_{ds}$-$V_{bg}$ characteristics at different bias ($V_{ds}$).*

2. **Low temperature transport characteristics and calculation of interfacial defect density from subthreshold swing ($SS$):**

The interfacial defect or trap density can be roughly determined from the estimated SS of a semiconductor device by using the equation

$$SS = \ln(10)\frac{K_B T}{e}\left(1 + \frac{e^2 D_{it}}{C_{total}}\right) \quad \text{...........................(S1.1)}$$

where, $C_{total}$ is the total capacitance which is basically a parallel combination of 285 nm SiO$_2$ and ~ 10 nm of hBN. $D_{it}$ is the defect density and $k_B$ is the Boltzmann constant. In room temperature (T ~ 295 K), SS is ~ 3.5 V/dec. This corresponds to $D_{it}$ ~ $3.8 \times 10^{11}$ cm$^{-2}$ eV$^{-1}$ which matches the calculated value using McWhorter carrier number fluctuation model.

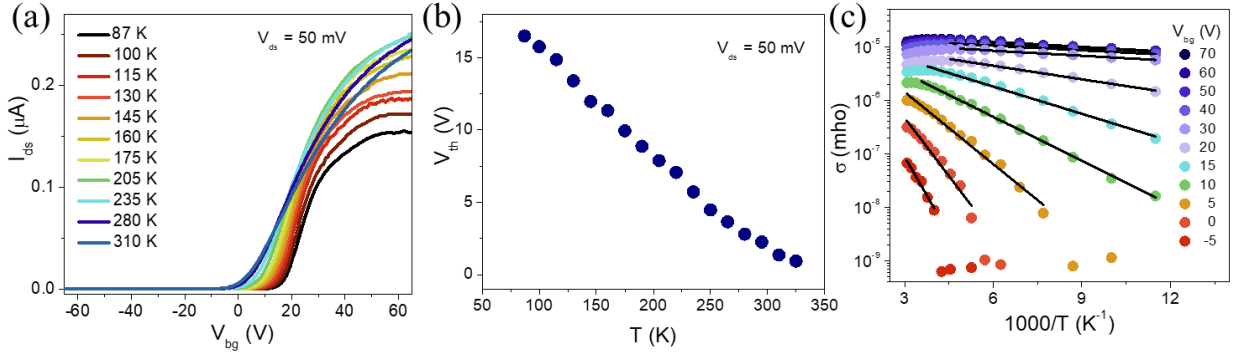

*Figure S2: Temperature dependent transport characteristics of the device: (a) Logarithmic plot of the transfer characteristics ($I_{ds}$-$V_{bg}$). (b) Shift of threshold voltage ($V_{th}$) as a function of temperature. (c) Conductivity as a function of temperature at different gate voltages.*

3. Low frequency noise measurement technique:

An *ac* two probe technique is used to measure the drain-source current and the current noise of the device. A lock-in amplifier (MFLI, Zurich Instruments) is used was used for phase-sensitive detection of the signal at a fixed reference frequency and phase. The lock-in technique significantly improves the measurement of low-noise signals by effectively rejecting contributions from frequencies and phases other than the reference, thereby enhancing the signal-to-noise ratio. Accurate and reproducible determination of the power spectral density (PSD) of conductance fluctuations is particularly challenging when the intrinsic sample noise is low, as instrumental noise often exhibits a 1/f dependence. To address this, the experimental setup was designed to minimize extrinsic noise contributions through a combination of hardware and software noise-reduction strategies. Software-based processing includes digitization, digital filtering, and fast Fourier transform (FFT) analysis, all implemented within a digital signal processing (DSP) framework. The use of advanced DSP algorithms and dedicated hardware further suppresses external interferences, enabling precise extraction of the intrinsic noise spectrum of the device. A schematic of the measurement setup and a representative current noise spectrum are shown in **Figure S3**.

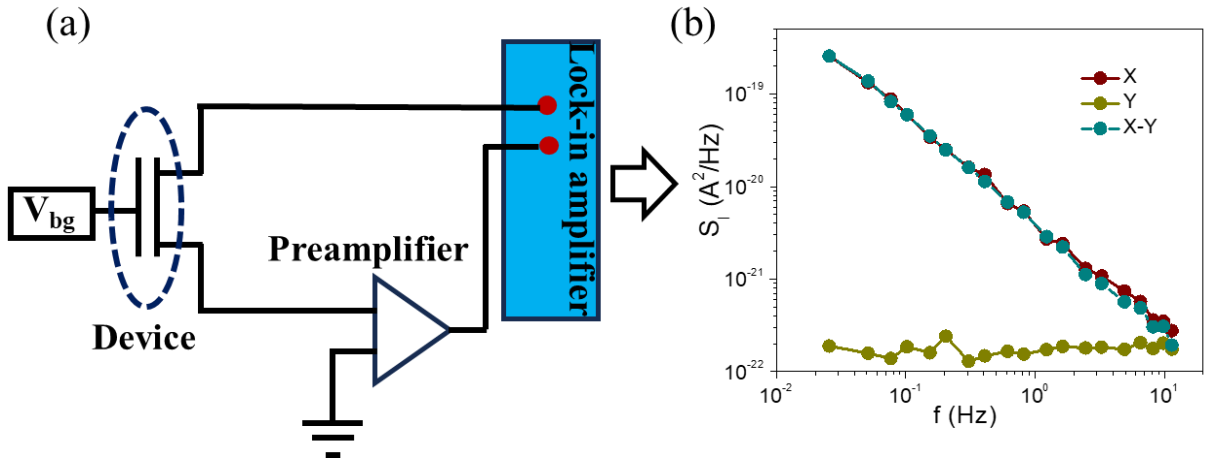

*Figure S3: Low frequency 1/f noise measurement technique: (a) Schematic diagram of the noise measurement setup. (b) Representative 1/f noise spectra of the device showing power spectral densities corresponding to X, Y and X-Y components of the time signal.*

4. **Drain-source bias ($V_{ds}$) dependent noise in room temperature:**

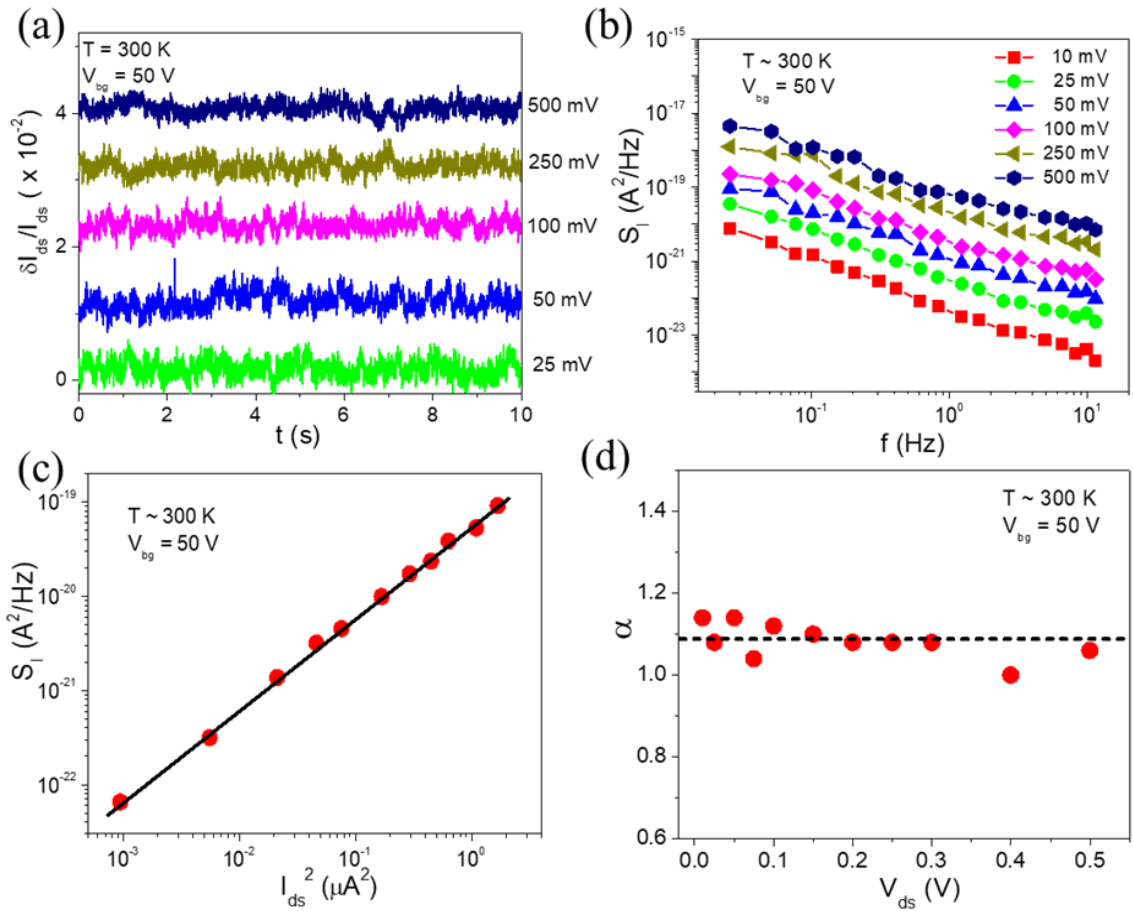

***Figure S4: Source drain bias ($V_{ds}$) dependent noise characteristics of the device:*** *(a) Time series data of normalized source-drain current ($I_{ds}$), and (b) the corresponding power spectral density versus frequency plots. (c) Integrated noise power spectral density ($S_I$), as a function of the square of source-drain current ($I_{ds}$). The black solid line indicates the linear fit to the data. (d) Noise exponent (α), as a function of source drain bias ($V_{ds}$). All the experiments are done under the same experimental conditions of T= 300 K, $V_{bg}$ = 50 V.*

## 5. Observation of $1/f$ and $1/f^2$ noise spectra:

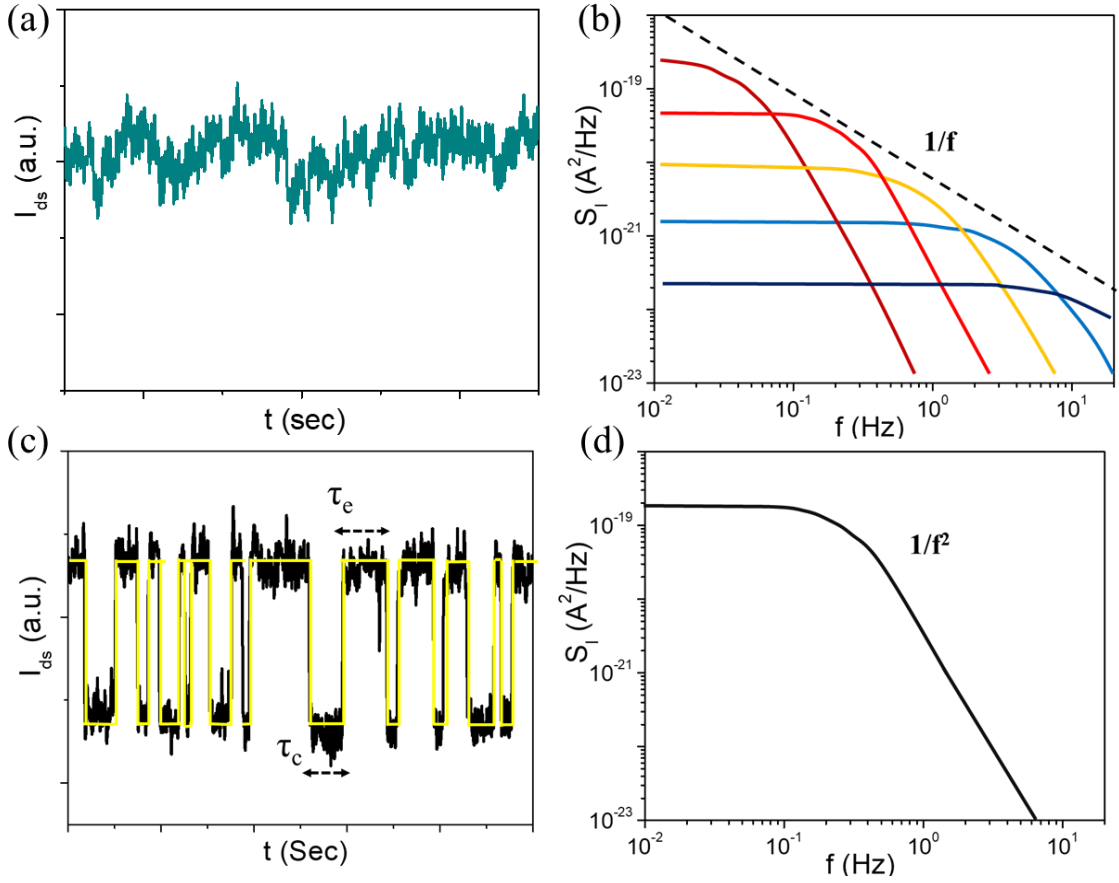

***Figure S5: Observation of $1/f$ and $1/f^2$ noise spectra:*** *(a) Time series data and (b) corresponding noise PSD of $I_{ds}$ with random fluctuations resulting from the superposition of several RTS. The noise PSD shows 1/f dependence and known as "flicker noise". (c) Time series data and (d) corresponding noise PSD of $I_{ds}$ with 'digital' two level fluctuations. The PSD shows Lorentzian ($1/f^2$) profile indicating one or few active defect level.*

## 6. Drain-source bias ($V_{ds}$) dependent noise in low temperature:

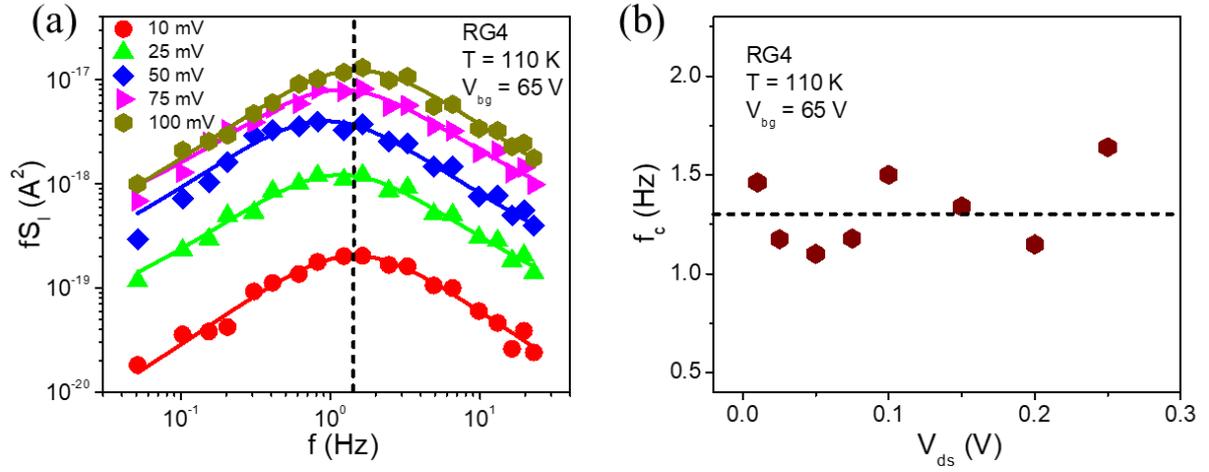

***Figure S6: Bias ($V_{ds}$) dependent noise PSD at low temperature:*** *(a) Plot of $fS_I(f)$ versus frequency f at different source drain biases $V_{ds}$. The solid line represented the Lorentzian fitting following the equation 1.4. Black dotted line indicates the positions of the Lorentzian peak ($f_c$) in the spectra. (b). Plot of $f_c$ as a function of $V_{ds}$ extracted from the PSD. The dotted line is a visual guide to the eye. All the experiments are done with same experimental conditions of T = 110 K, $V_{bg}$ = 65 V.*

## 7. Evaluation of time series, histogram and noise PSD of $I_{ds}$ at different gate voltages and temperatures:

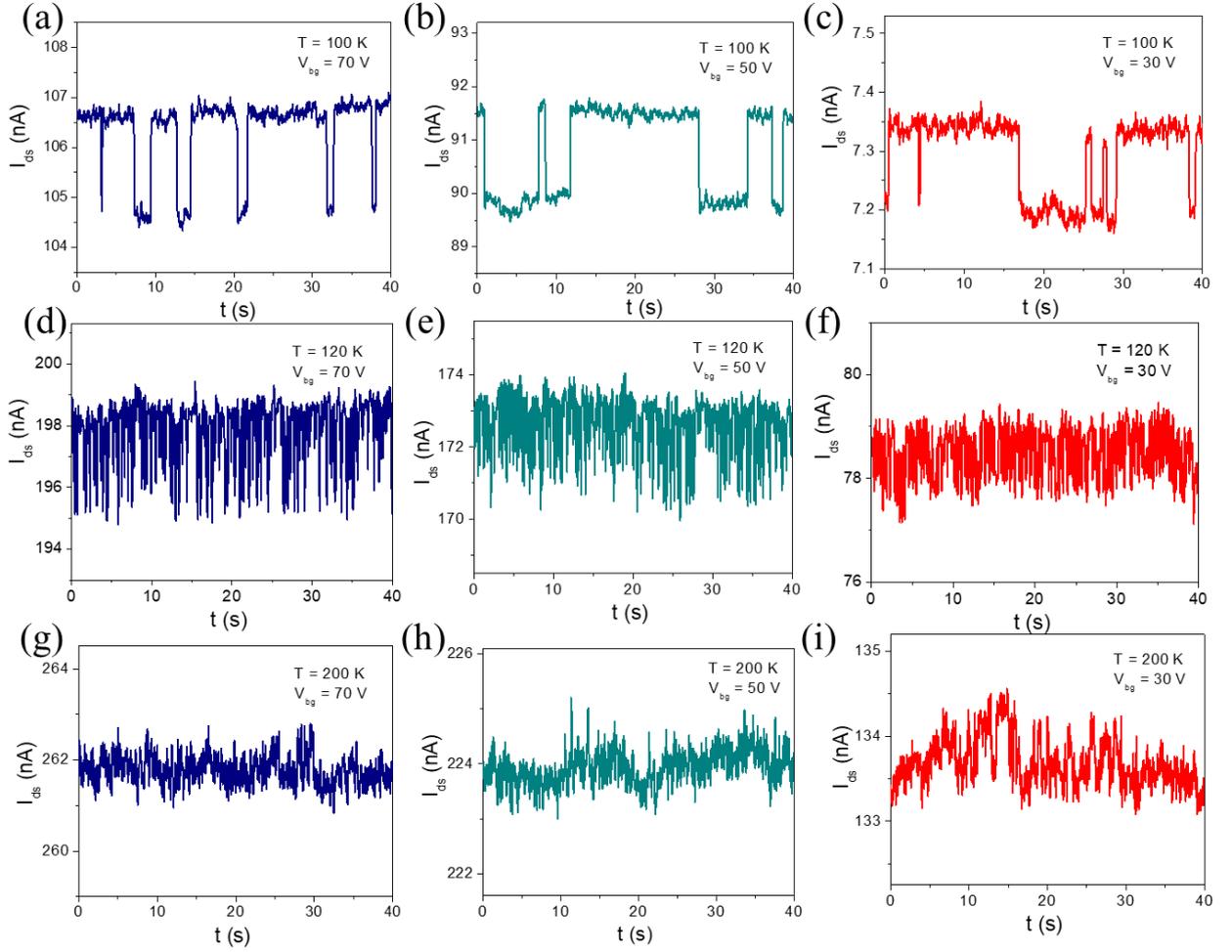

*Figure S7: Evaluation of time series data at different gate voltages and temperatures: (a)-(f) Various RTS traces observed in low temperatures. (g)-(i) RTS completely vanishes at the time series data at different gate voltages at T = 200 K.*

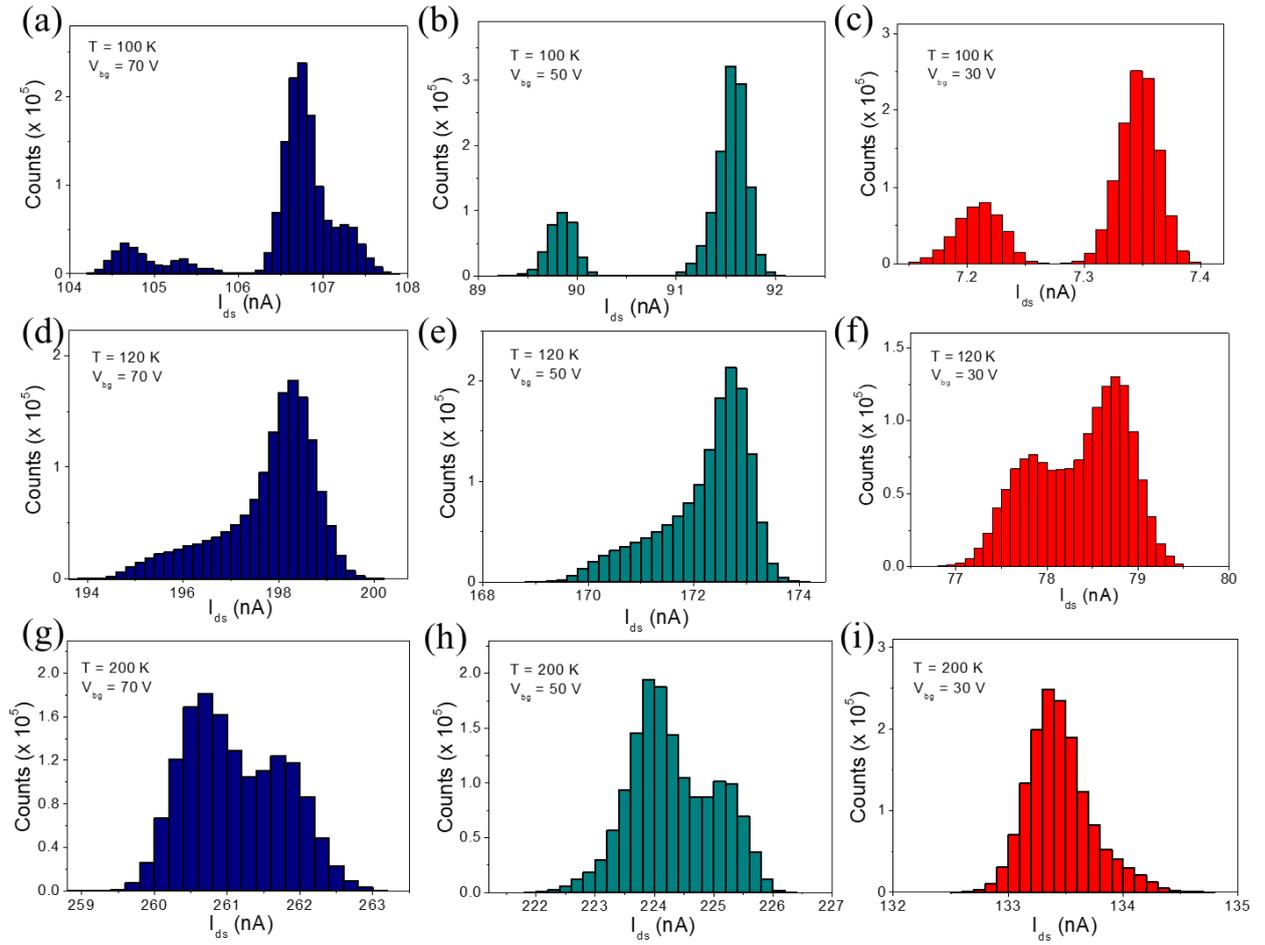

*Figure S8: Histogram plot of time series data (of Figure S7) at different gate voltages and temperatures.*

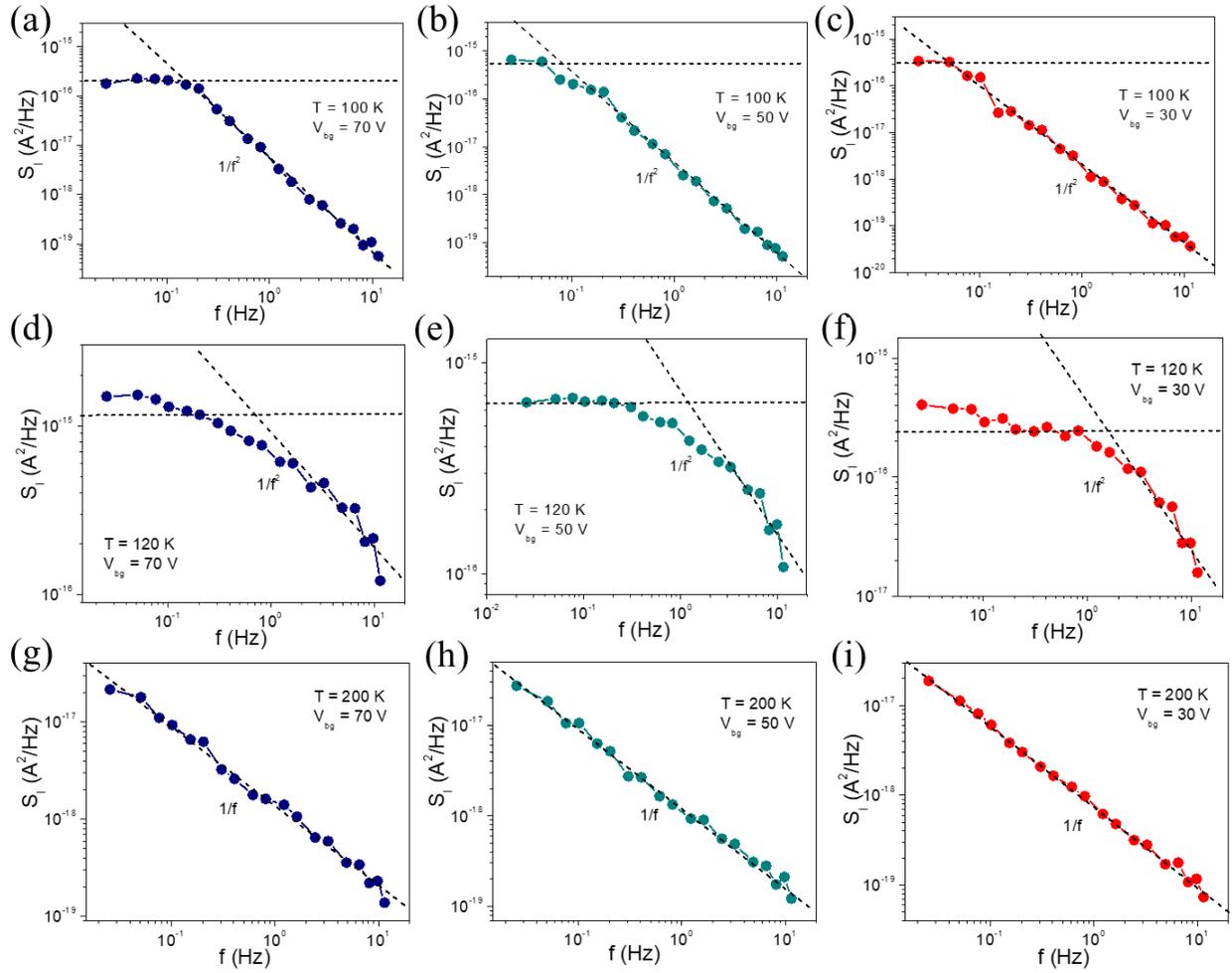

*Figure S9: Evaluation of low frequency noise PSD at different gate voltages and temperatures (of Figure S7): (a)-(f) At lower temperatures, the noise spectra follow a Lorentzian profile with a slope of $1/f^2$, suggesting the presence of RTS. (g)-(i) At T = 200 K, the noise spectra follow a flicker noise profile with a slope of $1/f$ indicating the absence of RTS.*

## 8. Evaluation of characteristics time constants at different gate voltages and temperatures:

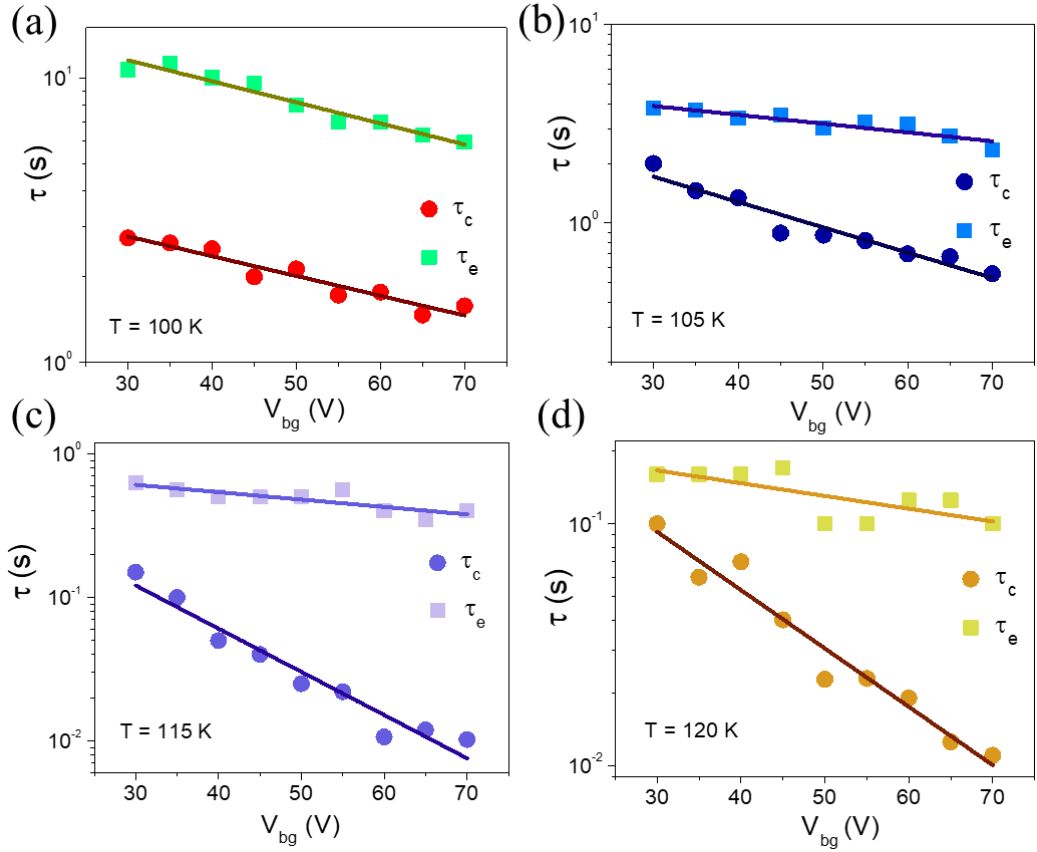

*Figure S10: Evaluation of characteristics time constants with gate voltages at different temperatures:* *(a) T = 100 K. (b) 105 K. (c) 115 K. (d) 120 K respectively.*

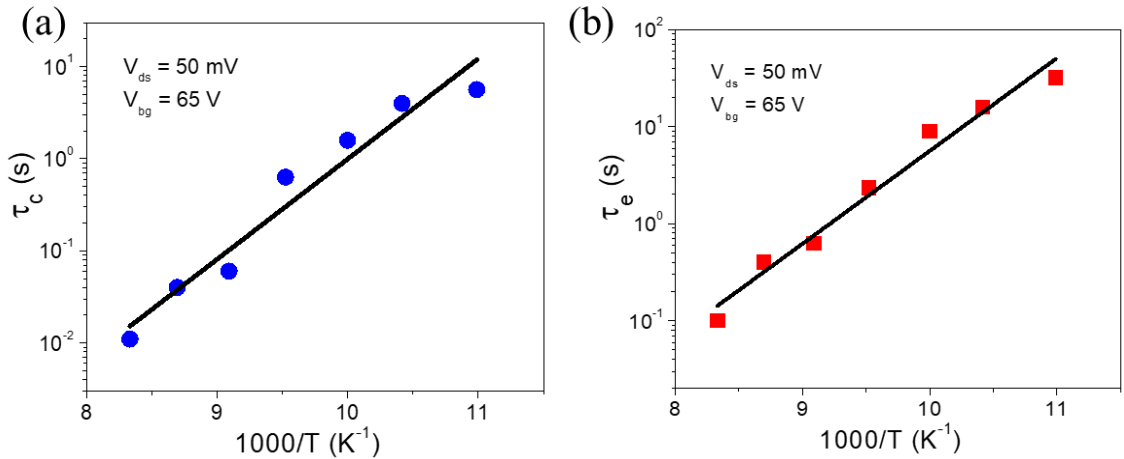

*Figure S11: Evaluation of characteristics time constants with temperatures:* *(a) $\tau_c$ (b) $\tau_e$ respectively. All are in the same experimental conditions of $V_{ds}$ = 50 mV, $V_{bg}$ = 65 V.*

## 9. Gate dependence and microscopic mechanism of low frequency noise:

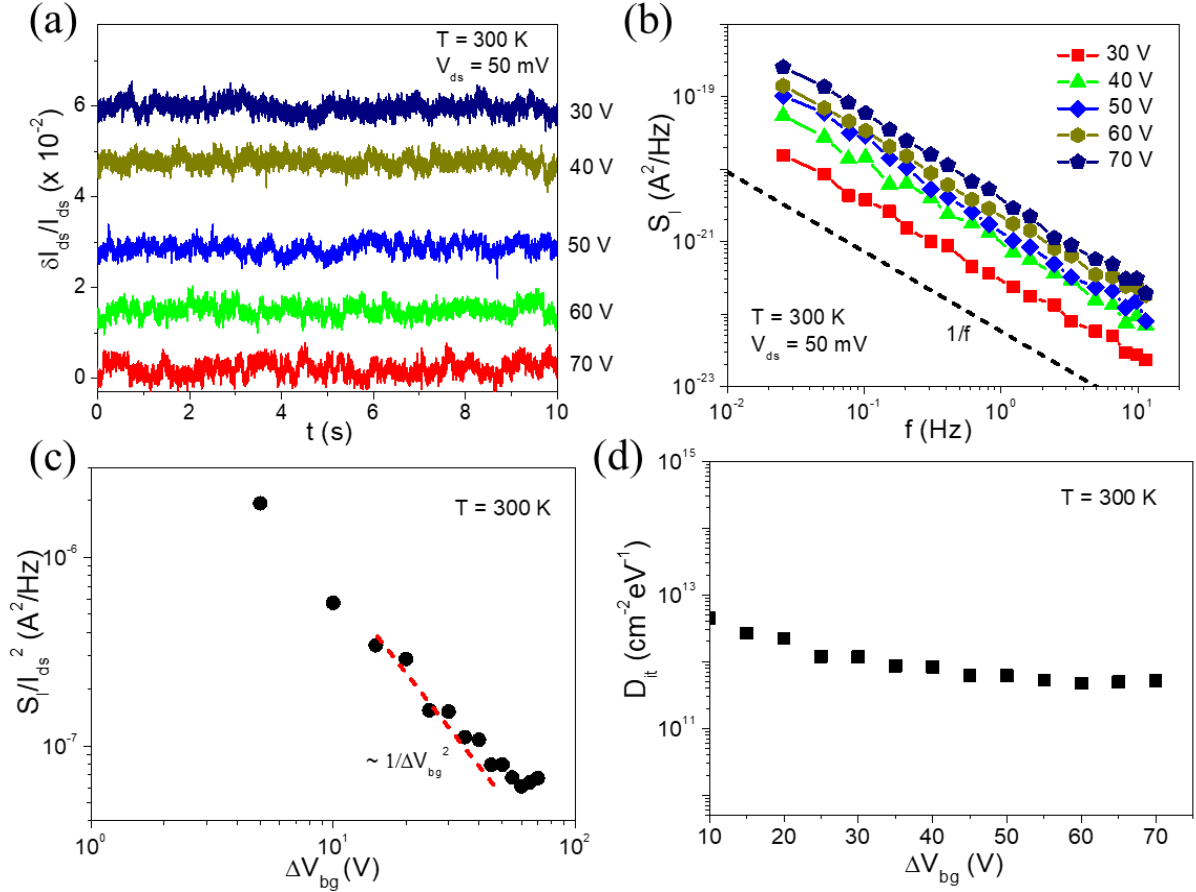

*Figure S12: Microscopic mechanism of low frequency noise:* (a) Time series data of source drain current ($I_{ds}$) with different back gate voltages ($V_{bg}$). (b) Gate voltage dependent low frequency noise at $V_{ds}$ = 50 mV. (c) Back gate dependent normalized $S_I/I_{ds}^2$ at room temperature. The dependence shows almost $1/\Delta V_{bg}^2$ at high density. (d) Calculated trap density ($D_{it}$) as a function of $\Delta V_{bg}$.

Electrical noise in TMDC systems has been attributed to a variety of factors, including Hooge mobility fluctuations (HMF)[1,2], McWhorter carrier number fluctuations (CNF)[3,4], and a mixture of the two as well[5,6]. To understand the mechanism of the low frequency noise, we performed and analysed carrier density dependent noise in our device. The parameter $\Delta V_{bg} = (V_{bg} - V_T)$ is considered to be approximately proportional to carrier density (*n*). **Figure S12a** and **S12b** represent the time series drain current fluctuation and low frequency noise at different gate voltages at room temperature ($T = 300\ K$) respectively. The noise PSD shows $1/f$ nature in the entire gate voltage range. In **Figure S12c**, the normalized noise PSD ($S_I/I_{ds}^2$) is represented as a function of $\Delta V_{bg}$. The low frequency noise behaviour of the device follows $S_I/I_{ds}^2 \propto 1/\Delta V_{bg}^2$, particularly at the higher gate voltages. This behaviour suggests the charrier number fluctuation in the semiconducting channel by correlating McWhorter model[7]. Following McWhorter model, the interfacial trap densities can be calculated by following the equation (1.8) (see **main manuscript**). The calculated trap density ($D_{it}$) ranges between ~ $\sim 4 \times 10^{11}\ cm^{-2}eV^{-1}$ and ~ $5 \times 10^{12}\ cm^{-2}eV^{-1}$ and shown in **Figure S12d**.

## 10. Low frequency $1/f$ noise in additional device (RG2) and comparison of noise figure of merit:

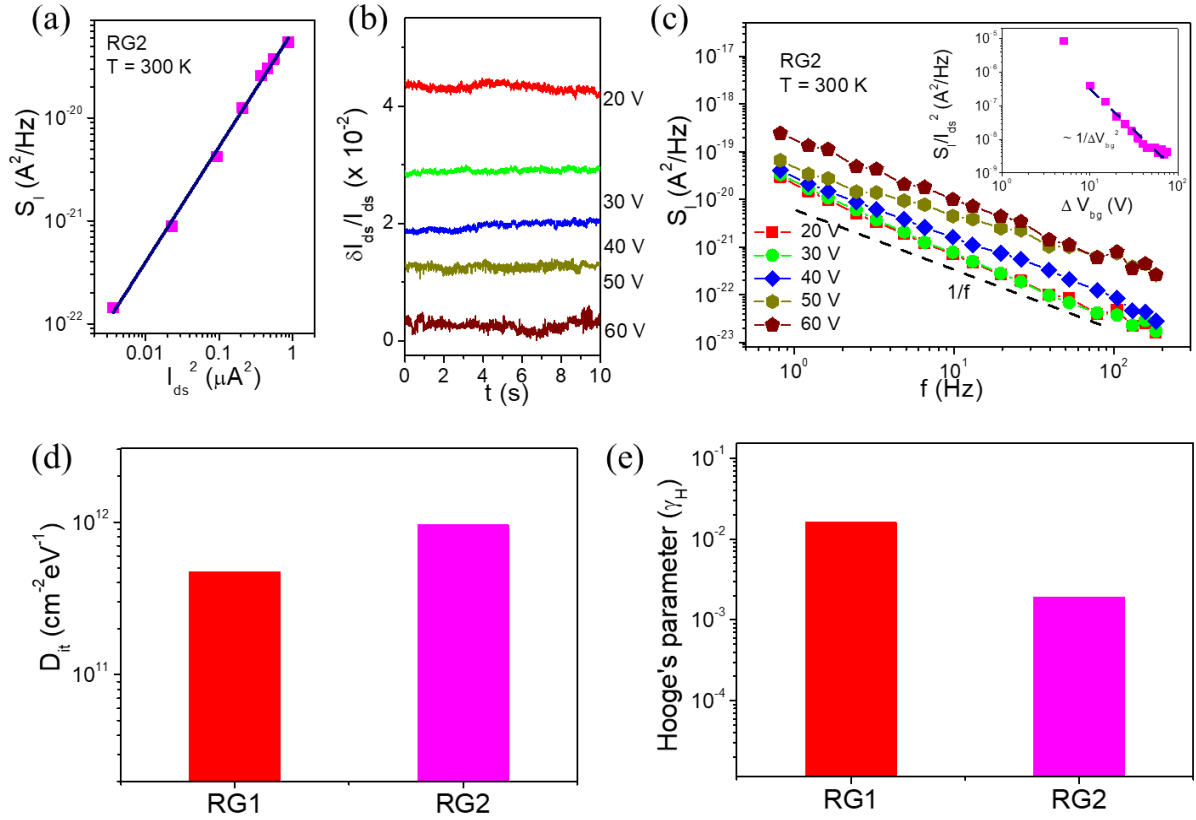

*Figure S13: Additional data for low frequency noise of an identical device RG1: (a) Noise power spectral density ($S_I$) as a function of square of source-drain current (Ids). (b) Variation of source drain current with back gate voltage ($V_{bg}$). (c) Gate voltage dependent low frequency noise. Inset shows the McWhorter fitting ($\sim 1/\Delta V_{bg}^2$) of normalized current noise ($S_I/I_{ds}^2$). All the data are taken in RG1 device at room temperature ($T = 300\ K$). Comparison of (d) trap density ($D_{it}$) and (e) Hooge's parameter of different fabricated devices at room temperature.*

To comprehend and compare the low frequency noise behaviour of the ReS$_2$ FET devices, more devices are fabricated and measured. RG1 is an identically fabricated device which shows similar linear ($I_{ds}^2$) dependence (**Figure S13a**). The time series drain current fluctuation and low frequency noise at different gate voltages at room temperature $(T = 300\,K)$ are represented by **Figure S13b** and **S13c** respectively. Similar to the RG FET, the fabricated RG1 FET follows McWhorter carrier number fluctuations (CNF) model $(S_I/I_{ds}^2 \propto 1/\Delta V_{bg}^2)$ also (**Figure S13c**, inset). **Figure S13d** compares the trap density of the devices. The calculated trap densities are almost similar ($10^{11}\,cm^{-2}eV^{-1}$) for both the devices.

The noise magnitude can be quantified using a dimensionless metric known as the Hooge parameter[8,9], which is frequently used to compare the level of noise among systems. The Hooge parameter is not affected by the channel length or width of FET devices, but rather by material-specific qualities such as material screening or coupling between substrate trap charges and channel charge carriers. This makes it an important metric for noise characterisation. The Hooge parameter can be defined using the phenomenological Hooge relation, which describes the power spectral density as:

$$\frac{S_I}{I_{ds}^2} = \frac{\gamma_H}{Nf^\alpha} \quad\quad\quad\quad\quad\quad\quad\quad\quad\quad\quad\quad\quad\quad (S1.2)$$

where, $\gamma_H$ is the phenomenological Hooge parameter, $N = C_{total}(V_{bg} - V_T)/q$ is the total number of free charge carriers in the sample and $\alpha$ is the exponent of noise.

Considering $\alpha \sim 1$ in the entire gate voltage range, the calculated Hooge parameter becomes $\sim 1.6 \times 10^{-2}$ for RG FET and $1.89 \times 10^{-3}$ for RG1 FET respectively at room temperature. The calculated Hooge parameters are comparable to values previously reported in literatures[1,10–12].

## 11. Observation of anamolous RTS:

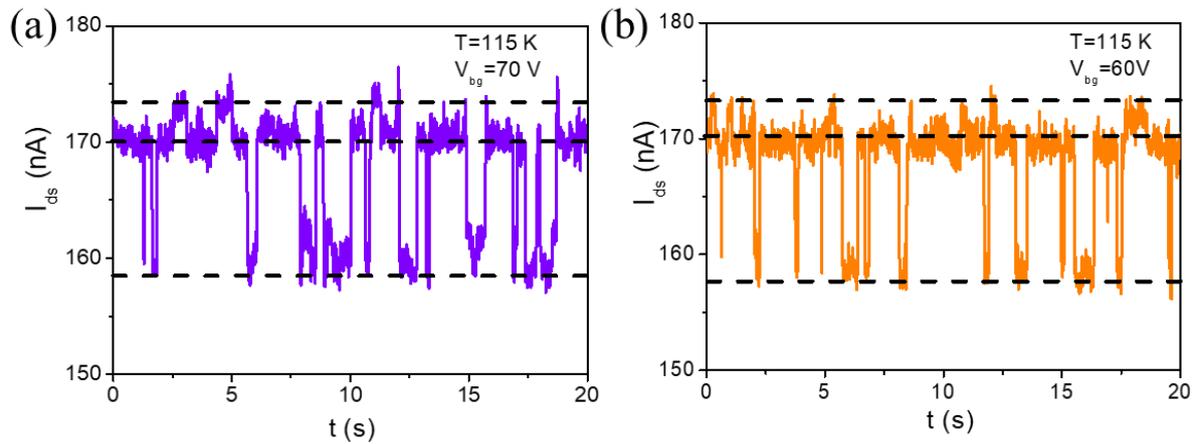

*Figure S14: Observation of anomalous RTS in ReS$_2$-hBN hybrid device at different gate voltages:* (a) $V_{bg}$ = 70 V. (b) $V_{bg}$ = 60 V.